\documentstyle[12pt]{article}
\def\bib{\bibitem}
\def\be{\begin{equation}}
\def\ee{\end{equation}}
\def\beqar{\begin{eqnarray}}
\def\eeqar{\end{eqnarray}}
\def\barr{\begin{array}}
\def\earr{\end{array}}
\def\lsim{\:\raisebox{-0.5ex}{$\stackrel{\textstyle<}{\sim}$}\:}

\def\and{\qquad {\rm and } \qquad}
\def\etal{ {\it et al.} }

\def\eg{ {\it e.g.} }

\def\etg{\eta_\gamma}
\def\etZ{\eta_Z}
\def\kg{\kappa_\gamma}
\def\kZ{\kappa_Z}
\def\lg{\lambda_\gamma}
\def\lZ{\lambda_Z}
\def\ka{\eta_a}
\def\kb{\eta_b}

\def\Pihat{\widehat \Pi}
\def\Shat{\widehat S}
\def\That{\widehat T}
\def\Uhat{\widehat U}
\def\Stil{\widetilde S}
\def\Ttil{\widetilde T}
\def\Util{\widetilde U}
\def\Pitil{\widetilde \Pi}
\def\la{\lambda_a}
\def\lb{\lambda_b}

\def\avk{\langle \eta \rangle}
\def\avkk{\langle \eta^2 \rangle}
\def\avl{\langle  \lambda \rangle}
\def\avll{\langle  \lambda^2 \rangle}
\def\avlk{\langle  \lambda \eta \rangle}
\def\gsw{$SU(2)_L \otimes U(1)_Y$ }
\def\prl#1{Phys. Rev. Lett. {\bf #1}}
\def\prd#1{Phys. Rev. {\bf D#1}}
\def\plb#1{Phys. Lett. {\bf B#1}}
\def\npb#1{Nucl. Phys. {\bf B#1}}

\begin{document}
\thispagestyle{empty}
\begin{flushright}
TIFR--TH/93--08 \\
May 1993\\
\end{flushright}

\vspace{5ex}
\begin{center}

{\large \bf Oblique electroweak corrections and triple vector boson
couplings}\\

\bigskip
\bigskip
{\sc
   Debajyoti Choudhury\footnote{debchou@tifrvax.bitnet\\
   Address after Sept.1, 1993: Max--Planck--Institut f\"ur
Physik, D--8000 M\"unchen 40, Germany.},
   Probir Roy\footnote{probir@tifrvax.bitnet}
   and
   Rahul Sinha\footnote{rahul@tifrvax.bitnet}
   }

\bigskip
{\it Tata Institute of Fundamental Research, Homi Bhabha Road,
Bombay 400 005, India.} \\

\vspace{5ex}
{\bf Abstract}\\

\medskip
\begin{quotation}
We relate all $C$-- and $P$--invariant anomalous triple
vector--boson couplings to the oblique electroweak parameters.
LEP constraints on the latter then yield the strongest and most
general simultaneous bounds to date on the former.  Even if the
oblique parameters assume  their Standard Model values
precisely, these bounds would not shrink to zero---thus
underscoring the need for direct experimental probes at future
colliders.

\end{quotation}
\end{center}
\newpage

 The \gsw  theory  of   electroweak interactions, which is a
part of the standard model (SM),  has had   dramatic
confirmations in  the  last  decade.  Experiments  at   LEP are
currently able to measure the  mass of the $Z$  and its
couplings   to fermions at a less than  $1\%$ level and they
agree  with the theory considered upto one loop.  This success
of the SM has  generated   a  feeling  that  even   those of
its parameters that have not been tested directly are likely to
be in good agreement with observation.  Such an impression  has
indeed  been fostered in the literature during the past  year
\cite{derujula} even with regard to forthcoming experiments such
as those planned at LEP 200.  While present precision measurements do
constrain    new physics  {\it vis-a-vis} the parameters of
the  SM, such constraints  need not be  as  restrictive in some
sectors as in others.

 In this Letter we examine how existing precision measurements
constrain the general $C$-- and $P$--conserving
triple--electroweak--vector--boson (TEVB) vertices
$WW\gamma$ and $WWZ$ which are predicted uniquely by the SM.
Much effort has already gone into obtaining such constraints
\cite{previous}, but there have   been two points   of
concern \cite{einhorn} involving some of  these calculations.
First, an \gsw gauge
non-invariant  Lagrangian has been used in extending the SM
gauge boson vertices. Second, the use  of  the cut--off procedure has
not been made with due care needed in the case of new physics.
In a  recent publication \cite{burgess}, Burgess and  London
have clarified these issues somewhat. They show that any
Lorentz and $U (1)_{em}$  gauge invariant Lagrangian, containing
$W's$ and  $Z's$, automatically
obeys \gsw gauge invariance, realized nonlinearly in general.
They have also pointed out misuses of the cutoff method
in estimating sizes of loop diagrams and have recommended
\cite{burgess} the use of dimensional regularization instead.

  In parameterizing new physics by an effective Lagrangian
approach, one can characterize the issue of decoupling versus
nondecoupling  in   the following way \cite{decoupling}. If a
gauge  invariant mass  parameter (\eg the  scale of new physics
$\Lambda$) in the underlying theory is large, the corresponding
particle decouples from  the  theory when the parameter tends to
infinity.  The decoupling  theorem requires that  the effective
Lagrangian have a renormalizable form apart from higher
dimensional terms containing inverse powers of $\Lambda$. If on
the other  hand, the   mass parameter increases on account of  a
dimensionless coupling constant becoming large, or if the mass
term is a gauge--variant one, then the particle does not
decouple.  This   is the case,   for example, with composite
models   of electroweak  vector  bosons, scenarios of
Technicolor or  with a heavy chiral fermion (such as the
top--quark) in a spontaneously broken theory.

  Restrictions on the TEVB couplings   in the
case  of a decoupled Lagrangian are being studied  by  other
authors \cite{zeppen}. These authors
have chosen to use a  Lagrangian  with an explicit linearly
realized \gsw gauge invariance.  This can be done when the
symmetry  breaking is  driven by elementary scalar fields.
We consider the more general case
\cite{burgess} of a nonlinear realization of gauge
invariance where the symmetry  breaking sector need not be
specified. Since a
Lagrangian with a nonlinearly realized gauge invariance is
equivalent \cite{burgess} to
one with $W's$   and $Z's$ and
satisfying $U(1)_{{\rm em}}$ gauge invariance, we use the latter
form to consider  extensions
to the TEVB interactions. This is tantamount to working with an
\gsw invariant Lagrangian in the
unitary gauge \cite{burgess,valencia}.

The $WWV \:(V=\gamma/Z)$ vertex, assuming $C$ and
$P$ invariance, can be parameterized by an effective Lagrangian
\cite{peccei}
\be
     {\cal L}_{\it eff}^{V}= -i g_{V}
             \left[ g_1^V
               \left( W^\dagger_{\alpha \beta} W^\alpha
                      - W^{\dagger\alpha} W_{\alpha \beta}
                \right) V^\beta
             +
               \kappa_V  W^\dagger_{\alpha} W_\beta
                               V^{\alpha\beta}
            + \frac{\lambda_V}{M_W^2}
                 W^\dagger_{\alpha \beta} {W^\beta}_\sigma
                 V^{\sigma\alpha} \right]
      \label{lagrangian}
\ee
Here $V_{\alpha\beta} = \partial_\alpha V_\beta - \partial_\beta
V_\alpha $, $W_{\alpha\beta} = \partial_\alpha W_\beta -
\partial_\beta W_\alpha $ and $g_{V}$ is the $WWV$ coupling
strength in the SM with $g_{\gamma}=e$   and   $g_{Z}=ec/s$,
where $c^2 \equiv 1 -s^2 \equiv M_W^{2}/M_Z^2$.     The  SM
values for the extra new couplings are $g_1^Z= \kg =\kZ =1$, $
\lg=\lZ=0$. Electromagnetic  gauge invariance fixes $
g_1^\gamma$ to be unity.  The other couplings
$g_1^Z,\kg,\kZ,\lg,\lZ$ have to be determined  experimentally.
The number of extra parameters is further reduced by taking
\cite{g1z} $g_1^Z$, the weak neutral charge of the $W$, equal to
unity. The Lagrangian in (\ref{lagrangian}) has  the advantage
that  it  could represent either decoupling or nondecoupling new
physics at high energies and provides a practical way of
calculating  two  point functions of  physical vector bosons.

New physics beyond the SM can be usefully constrained  in  terms
of  the  ``oblique'' electroweak
\cite{peskin,altabarb} parameters $\Stil, \Ttil,$ and $\Util$ (
or $\Delta \epsilon_1, \Delta \epsilon_2,$ or $\Delta
\epsilon_3$ ).  These are linearly related  \cite{ftnote} to
$\Pitil_{WW}(q^2),\: \Pitil_{ab}(q^2)$ (where  $a,b =\gamma,Z$),
which are the new physics contributions to the electroweak
vector boson self--energy functions. Several determinations
\cite{ellis} have been made on these parameters using LEP
data as well as lower energy information. If the anomalous
TEVB couplings constitute the sole source of new physics,
$\Ttil$ and $\Util$ will measure the weak isospin breaking
induced by them, while $\Stil$ arises from their \gsw breaking
aspect through the involvement of the longitudnal vector boson
modes.

One can work out the divergent contributions to the
$\Pitil$--functions from the general $WWV$ vertex by
evaluating the 1--loop graphs with virtual vector bosons using
(\ref{lagrangian}) and dimensional regularization.  (Anomalous
four vector boson vertices, proportional to $\lambda_V$, make
vanishing contributions to the seagull
loops  and hence to the $\Pitil$--functions.)
With $\epsilon =
4 - $ dimension of spacetime and $\mu \equiv $ mass regulator, we can
write
\be
\Pitil (q^2) =
\Pihat (q^2) \: \left[
2/\epsilon - \gamma_E +
       \ln \left( \frac{4 \pi \mu^2}{M_Z^2} \right)
       \right]
       + \cdots
        \label{pihat}
\ee
where the dots denote finite terms which are small for
$\left| q^2 \right| \lsim M_Z^2$. We compute
\be
\barr{rl}
\displaystyle \Pihat_{ab} (q^2) =
 - \frac{g_{a} g_{b} q^2}{192 \pi^2}
& \displaystyle
\left[
    \left( \ka + \kb \right) \left( 36 - 4 r - r^2 \right)
    + \ka \kb  r \left( 2 - r \right)
  \right. \\
& \displaystyle
    \left.
     + \left( 3 \la + 3 \lb + \ka \lb + \la \kb \right)
          \left( 24 - 4 r \right)
     + \la \lb  \left( 36 + 8 r - 2 r^2 \right)
      \right],
\earr
          \label{pineut}
\ee
where $r \equiv q^2/M_W^2$ and $\eta_{a,b} \equiv 1 - \kappa_{a,b}$.
For compactness, define
\[
\barr{rclcrcl}
{\cal C}_1 &\equiv& 36 c^2 - 4 - c^{-2},
  &\qquad&
{\cal C}_2 &\equiv& 24 c^2 - 4,\\
  {\cal C}_3 &\equiv& 2 - c^{-2},
  &\qquad&
  {\cal C}_4 &\equiv& 36 c^2 + 8 - 2 c^{-2} \:,\\
\earr
\]
With $\That$, $\Shat$
and $\Uhat$  defined analogously \cite{ftnote} in terms of the
$\Pihat$'s and denoting $\langle F
\rangle \equiv s^2 F_\gamma + c^2 F_Z$ for any $F$, we obtain
\be
\displaystyle \Shat =
- \frac{1}{12 \pi}
\left[
  (\etZ - \etg) \left\{ {\cal C}_1 + \avk {\cal C}_3
                      + \avl {\cal C}_2
              \right\}
      +
  (\lZ - \lg) \left\{ {\cal C}_2 \left(3 + \avk \right)
                      + \avl {\cal C}_4
              \right\}
       \right].
\ee
Since $\Pihat_{ab}(0) =0$, only $\Pihat_{WW}(0)$ contributes to
$\That$ and we can write
\be
\displaystyle
\Pihat_{WW} (q^2) =
 \alpha M_W^2\left[ \That - \frac{r}{48 \pi s^2} \: {\cal R}(r)
             \right]\:,
    \label{piww}
\ee
\be
\barr{rcl}
\displaystyle
\That & = &
\displaystyle
- \frac{3}{16 \pi}
\left[ \left\{ 4 + \etg \right\} \etg +
       \frac{\etZ}{c^2 s^2}
           \left\{ 2 \left( 2 c^4 + 2 c^2 - 1 \right)
                   + \left( c^4 + c^2 - 1 \right) \etZ
           \right\}
\right]\:,
  \\[3ex]
\displaystyle
{\cal R}(r)   & \equiv &
\displaystyle
      \left( 4 - 2 r \right)
      \left\{ 5 \avk + \avkk \right\} + 28 \etZ + 5 \etZ^2
      + 8 \left( 3 -r \right) \left\{ 3 \avl + \avlk \right\}\\
& & \displaystyle
  + 24 \lZ \left\{ 3 + \etZ \right\}
  + 2 \left( 6 + 2 r -  r^2 \right) \avll
  + 4 \left( 3 + 3 c^{-2} + r \right) \lZ^2
\: .
\earr
         \label{piww-def}
\ee
Consequently,
\be
 \displaystyle
 \Uhat =
  - \frac{1}{12 \pi} {\cal R} (1)
  +     \frac{1}{12 \pi c^2}
 \left[ 2 {\cal C}_1 \avk +  {\cal C}_3 \avk^2
        + 2 {\cal C}_2 \avl \left( 3 + \avk \right)
        +  {\cal C}_4 \avl^2 \right] \: .
\ee
Since $\That$ depends only on $\Pihat_{WW}(0)$ and
$\Pihat_{ZZ}(0)$ and not their $q^2$ variations, it is unaffected
by the dimension 6
operators and is hence independent of $\lg$ and $\lZ$.  Also,
terms in $\Shat$ are proportional to either $\etZ-\etg$ or
$\lZ-\lg$ as they should, since $\Shat$ originates from the
mixing between weak hypercharge ($Y$) and the third component of
weak isospin and the $WWY$ vertex is linear in these
differences.

The oblique
parameters are not finite quantities here owing to
 a nonrenormalizable Lagrangian.  In a
cut--off dependent regularization scheme, this fact would
manifest itself through a non--trivial functional dependence on
the cut--off scale
\cite{zeppen}. As a matching condition between two effective theories
\cite{georgi},
we identify $\mu = \Lambda$, the scale at which new physics becomes
manifest (assumed to be $\sim 1\:TeV$). Using the $\overline{\rm MS}$
scheme of renormalization, we can then write (see eqn.\ref{pihat})
\be
\displaystyle
\Stil \simeq  \Shat\:   \ln \frac{\Lambda^2}{M_Z^2}
\ee
where we have retained only the largest logarithms.
Similar relations hold for $\Ttil$ and $\Util$.
Observed bounds for $\Stil$, $\Ttil$ and $\Util$ can now be
translated onto $\Shat$, $\That$, $\Uhat$.

We use \cite{ellis} $\Stil = -0.31 \pm 0.49$, $\Ttil = -0.12 \pm
0.34$ and $\Util = -0.11 \pm 0.92$, though our results are
insensitive to the central values.  $\Ttil$ allows  only
an elliptic band
in the $\kg - \kZ$ plane (Fig. $1 a$), where the width is given
by the errors (95\% C.L.) on $M_W/M_Z$ and $\Ttil$.   $\Stil$ and $\Util$
then reduce the
allowed region to only the small shaded part of the elliptic
band. This is shown enlarged in Fig.($1 b$).
Since $\Stil$ is proportional to the differences of
$\gamma$-- and $Z$--couplings, constraints on it generally
(though not always) tend to make those converge.  Rather
unexpectedly, $\Util$ plays a significant role in constraining
these anomalous couplings. In conjunction with $\Stil$ and
$\Ttil$, it serves to exclude a large part of the parameter
space. We consider the $\kg - \lg$ plane (Fig. 2),  since
comparison with direct observations at {\sl UA2} and $CDF$ is
then possible. Also shown is part of an ellipse, the interior of
which represents (at $95\%$ C.L.) the area allowed by the {\sl
UA2} data \cite{UA2}. Moreover, the two parentheses on the
$\lg$--axis indicate the region in $\etg$ allowed by the $CDF$
data \cite{rahul} assuming $\lg = 0$.  Of course the constraints
are equally tight when expressed in any other form, say in the
$\kZ $--$\lZ$ plane, as shown in Fig. 3.

These constraints on TEVB vertices are stronger than those
achieved so far by direct experiments. Whereas we obtain the ($95\%$
C.L.) bounds $-6.1 \lsim
\kg - 1\lsim 4.1$, $-6.0\lsim \lg \lsim 4.5$, $-2.0 \lsim \kZ - 1
\lsim 0.3$ and $-4.5 \lsim \lZ \lsim 1.9$, earlier
{\sl UA2} analysis \cite{UA2} had yielded $ -8.4 \lsim \kg-1 \lsim
12.1 $ (for arbitrary $\lg$) and $-8.5
\lsim \lg \lsim 6.5$ (for arbitrary $\kg$).
The allowed regions represent solutions to polynomial equations,
which are  curves that thicken into bands on account of experimental
error bars on the coefficients.  As the errors shrink to zero,
the allowed parameter space collapses into the curves still
permitting wide ranges of values for the anomalous couplings,
\eg the solid curves in  Figs.  ($1b,\:2,\:3$) correspond to the
SM point {\it viz.} $\Stil = \Ttil = \Util = 0$. (This is  a
consequence of cancellations between various contributions to
the $\Pitil$--functions.) Direct studies of the TEVB vertices at
Fermilab, LEP 200 or the NLC will be necesssary to probe regions
much closer to the origin \cite{baur}.

To conclude, our use of precision measurements at LEP and at
lower energies, in terms of $\Stil, \Ttil$ and $\Util$,
constrains the anomalous $WW\gamma$ and $WWZ$ vertices quite
stringently.  Unlike previous efforts, which could constrain
only $\kg$ and $\lg$, we are able to restrict $\kZ$ and $\lZ$ as
well. These bounds are much stronger than the all existing
limits, though comparable limits may be directly achieved at
Fermilab in the near future.  The three oblique parameters do
constrain the $WW\gamma$ and $WWZ$ anomalous couplings, but
cannot exclude all regions in the space of the four coupling
constants --- there is always a set of limiting curves.
Stronger restrictions cannot be imposed from these measurements
alone; direct experimental study of TEVB vertices at future
colliders would be necessary to go beyond. This point of
contention has been resolved by the present work.

We thank S.~Banerjee,  C.P.~Burgess and A.~Gurtu for
discussions.
\newpage
\centerline{\bf Figure Captions}

\begin{enumerate}
\item
($a$) The dashed elliptical band gives the constraints from $\Ttil$ alone on
the $\kg$--$\kZ$ plane for $\Lambda = 1\;TeV$.
In this and the following figures, the shaded part indicates the
region  allowed by $\Stil$, $\Ttil$ and $\Util$ constraints put together.
($b$) Enlarged section of the shaded area in ($a$). In this and
subsequent figures, the solid
curve represents solutions for $\Stil =\Ttil = \Util =0$.
\item
Constraints (at 95\% C.L.) from $\Stil$,$\Ttil$ and $\Util$ on
the $\kg$--$\lg$ plane for $\Lambda = 1\;TeV$. {\sl UA2} data
constrain the parameters to be bound by the dashed ellipse.
The parentheses on the $\lg$ axis give the bounds of Ref.
\protect\cite{rahul}
\item
Constraints (at 95\% C.L.) from $\Stil$,$\Ttil$ and $\Util$ on
the $\kZ$--$\lZ$ plane for $\Lambda = 1\;TeV$.
\end{enumerate}

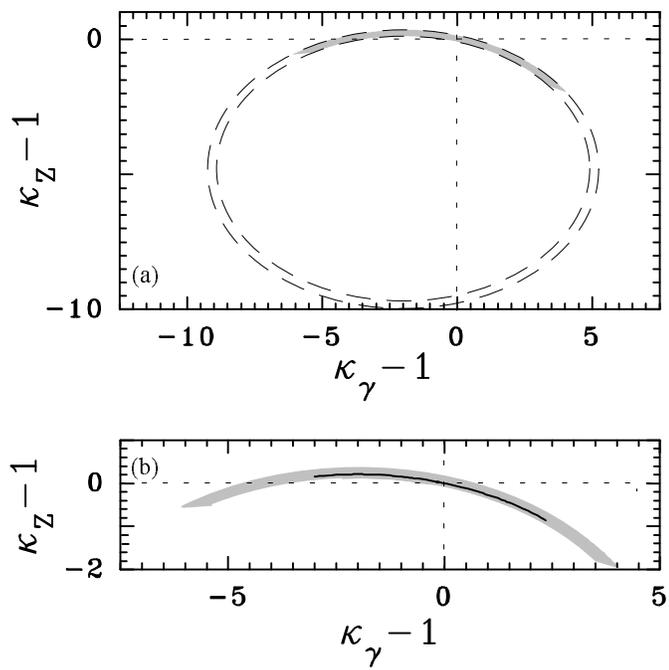
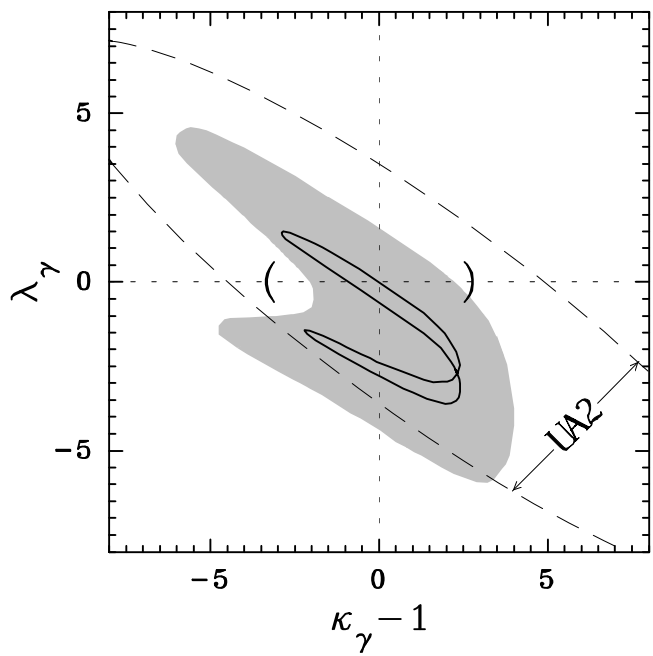
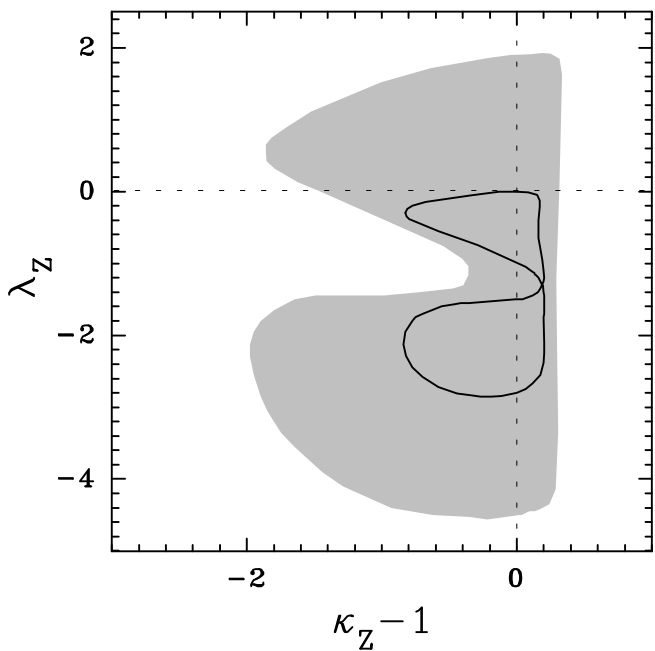
\end{document}